\documentclass[aps,prl,showpacs,english,10pt,a4,nofootinbib,notitlepage,twocolumn,superscriptaddress]{revtex4}

\usepackage[utf8]{inputenc}
\usepackage[T1]{fontenc}
\usepackage[english]{babel}

\usepackage{times}
\usepackage{dsfont}
\usepackage{multirow}
\usepackage{amssymb,amsmath,amsfonts,amsthm}
\usepackage{mathtools}
\usepackage{verbatim}
\usepackage{enumerate}
\usepackage{graphicx}
\graphicspath{{figs/}}
\usepackage{hyperref}
\usepackage{bbm}
\usepackage{booktabs}

\usepackage{pdfpages}

\usepackage[caption=false]{subfig}

\usepackage{color}
\definecolor{dred}{rgb}{.8,0.2,.2}
\definecolor{ddred}{rgb}{.8,0.5,.5}
\definecolor{dblue}{rgb}{.2,0.2,.8}
\definecolor{dgreen}{rgb}{.2,0.5,.2}

\newcommand{\bra}[1]{\mbox{$\langle #1|$}}
\newcommand{\ket}[1]{\ensuremath{|#1\rangle}}

\newcommand{\be}{\begin{equation}}
\newcommand{\ee}{\end{equation}}

\newcommand{\bea}{\begin{eqnarray}}
\newcommand{\eea}{\end{eqnarray}}

\begin{document}

\title{A Quantum Algorithm for Solving Linear Differential Equations: Theory and Experiment}
\date{\today}
\author{Tao Xin}
\affiliation{State Key Laboratory of Low-Dimensional Quantum Physics and Department of Physics, Tsinghua University, Beijing 100084, China}
\affiliation{Shenzhen Institute for Quantum Science and Engineering and Department of Physics, Southern University of Science and Technology, Shenzhen 518055, China}
\affiliation{Tsinghua National Laboratory of Information Science and Technology and The Innovative Center of Quantum Matter,  Beijing 100084, China}

\author{Shijie Wei}
\affiliation{State Key Laboratory of Low-Dimensional Quantum Physics and Department of Physics, Tsinghua University, Beijing 100084, China}
\affiliation{IBM research, Beijing 100094, China}

\author{Jianlian Cui}
\affiliation{Department of mathematics, Tsinghua University, Beijing 100084, China}

\author{Junxiang Xiao}
\affiliation{State Key Laboratory of Low-Dimensional Quantum Physics and Department of Physics, Tsinghua University, Beijing 100084, China}

\author{Iñigo Arrazola}
\affiliation{Department of Physical Chemistry, University of the Basque Country UPV/EHU, Apartado 644, 48080 Bilbao, Spain}

\author{Lucas Lamata}
\affiliation{Department of Physical Chemistry, University of the Basque Country UPV/EHU, Apartado 644, 48080 Bilbao, Spain}

\author{Xiangyu Kong}
\affiliation{State Key Laboratory of Low-Dimensional Quantum Physics and Department of Physics, Tsinghua University, Beijing 100084, China}

\author{Dawei Lu}
\email{ludw@sustc.edu.cn}
\affiliation{Shenzhen Institute for Quantum Science and Engineering and Department of Physics, Southern University of Science and Technology, Shenzhen 518055, China}

\author{Enrique Solano}
\affiliation{Department of Physical Chemistry, University of the Basque Country UPV/EHU, Apartado 644, 48080 Bilbao, Spain}
\affiliation{IKERBASQUE,  Basque  Foundation  for  Science,  Maria  Diaz  de  Haro  3,  48013  Bilbao,  Spain}
\affiliation{Department of Physics, Shanghai University, 200444 Shanghai, China}

\author{Guilu Long}
\email{gllong@tsinghua.edu.cn}
\affiliation{State Key Laboratory of Low-Dimensional Quantum Physics and Department of Physics, Tsinghua University, Beijing 100084, China}
\affiliation{Tsinghua National Laboratory of Information Science and Technology and The Innovative Center of Quantum Matter,  Beijing 100084, China}

\begin{abstract}
We present and experimentally realize a quantum algorithm for efficiently solving the following problem: given an $N\times N$ matrix $\mathcal{M}$, an $N$-dimensional vector $\textbf{\emph{b}}$, and an initial vector $\textbf{\emph{x}}(0)$, obtain a target vector $\textbf{\emph{x}}(t)$ as a function of time $t$ according to the constraint $d\textbf{\emph{x}}(t)/dt=\mathcal{M}\textbf{\emph{x}}(t)+\textbf{\emph{b}}$. We show that our algorithm exhibits an exponential speedup over its classical counterpart in certain circumstances. In addition, we demonstrate our quantum algorithm for a $4\times4$ linear differential equation using a 4-qubit nuclear magnetic resonance quantum information processor. Our algorithm provides a key technique for solving many important problems which rely on the solutions to linear differential equations.
\end{abstract}

\maketitle

\textit{Introduction. }--
Linear differential equations (LDEs) are an important framework with which to describe the dynamics of a plethora of physical models, involving classical as well as quantum systems. They are playing key roles in many applications, e.g., predicting climate change and calculating fusion
energy. In fact, many of the main applications of supercomputers are in the form of large systems of differential equations \cite{Kothe09}. Generally, solving an LDE is a hard problem for a classical high-performance computer, in particular when the size of the configuration space is large, as for example in quantum systems or fluid dynamics.

A possible way to overcome the above difficulty is to utilize quantum computing. Quantum information processing is one of the most fruitful fields of research in physics nowadays. Besides the famous Shor's factoring algorithm \cite{Shor97,Simon97} and Grover's search algorithm \cite{Grover97} , a quantum computer is also capable of solving linear systems of equations~\cite{Harrow09,Subasi18} exponentially faster than any classical computers. In recent years, first steps towards solving linear equations have been demonstrated in optics \cite{Barz14,Cai14}, nuclear magnetic resonance (NMR) \cite{Pan14, konglinear}, and superconducting circuits \cite{sclinear}. However, extending the algorithm to differential equations is not straightforward. Although some quantum algorithms have been proposed~\cite{Leyton08,Berry14,Berry17}, they are not easily implemented using state-of-the-art techniques due to the lack of quantum circuits. Therefore, it is timely to design an implementable quantum algorithm  and carry out first experimental demonstrations for solving LDEs in controllable quantum platforms.

Here, we present a quantum algorithm for solving LDEs only comprising of universal set of quantum gates. The precision of our algorithm can be boosted exponentially by adding the number of ancilla qubits. We further demonstrate it in a 4-qubit NMR system, which is a quantum platform with a myriad of successes in the field of quantum technologies \cite{nmrtao}. Many of the first demonstrations of quantum algorithms were achieved in this platform \cite{grover1,grover2,shor1,shor2,DJ1,order1,mosim,holo}, which inherits the high-degree of quantum control in NMR spectroscopy during the twentieth century. This includes the recent demonstration of quantum machine learning \cite{sim69} and linear solvers of equations \cite{Pan14}. In this work, we carry out a proof-of-principle experiment to implement an LDE solver in a 4-qubit NMR quantum processor.

\textit{Problem. }--
Here is the description of the problem for solving LDEs. An unknown vector $\textbf{\emph{x}}(t)$ starts from an initial point $\textbf{\emph{x}}(0)$ and follows an evolution described by an LDE $d\textbf{\emph{x}}(t)/dt=\mathcal{M}\textbf{\emph{x}}(t)+\textbf{\emph{b}}$, where $ \mathcal{M} $ is an arbitrary $ N \times N $ matrix, while $\textbf{\emph{b}}$ and $\textbf{\emph{x}}(t)$ are $N$-dimensional vectors.

\textit{Algorithm. }--
The analytical solution of the equation can be written as $\textbf{\emph{x}}(t)=e^{\mathcal{M}t}\textbf{\emph{x}}(0)+(e^{\mathcal{M}t}-I)\mathcal{M}^{-1}\textbf{\emph{b}}$. If the exponential evolution $e^{\mathcal{M}t}$ and the inverse operator $\mathcal{M}^{-1}$ can be effectively realized, one can easily obtain the solution $\textbf{\emph{x}}(t)$. In the following, we present the basic idea of finding $\textbf{\emph{x}}(t)$ based on a quantum algorithm. By Taylor expansion,
the solution $\textbf{\emph{x}}(t)$ is approximately
\begin{eqnarray}
\textbf{\emph{x}}(t) \approx  \sum_{m=0}^{k}\dfrac{(\mathcal{M}t)^{m}}{m!}\textbf{\emph{x}}(0)  +\sum_{n=1}^{k}\dfrac{\mathcal{M}^{n-1}t^{n}}{n!}\textbf{\emph{b}},
\end{eqnarray}
where $k$ is the approximation order. Vectors $\textbf{\emph{x}}(0)$ and $\textbf{\emph{b}}$ can be described by quantum states $\ket{x(0)}=\sum_j x_{j}(0)/\|x(0)\|\ket{j}$ and $\ket{b}=\sum_j b_{j}/\|b\|\ket{j}$, respectively, where $x_j(0)$ and $b_j$ are the $j$-th elements of these vectors, $\ket{j}$ is the $N$-dimensional computational basis, and $\| \cdot \|$ is the module operation. Matrix $\mathcal{M}$ can be described by operator $A$ defined as $A=\sum_{i,j}\mathcal{M}_{ij}/\| \mathcal{M} \| \ket{i}\bra{j}$. Hence, the $k$-th order approximate solution converts to
\begin{align}
\label{sim}
\ket{x(t)} \approx & \sum_{m=0}^{k}\dfrac{\| x(0)\|(\| \mathcal{M} \|At)^{m}}{m!}|x(0)\rangle  \\\nonumber
 & +\sum_{n=1}^{k}\dfrac{\| b \| (\| \mathcal{M} \| A)^{n-1}t^{n}}{n!}|b\rangle
\end{align}
up to normalization. Our algorithm provides a general framework for computing Eq. (\ref{sim}) employing a quantum system with the assistance of ancilla qubits. The algorithm works for both unitary and non-unitary $A$'s, and in the following we consider each of the two situations, respectively.

\textit{Case I}: If operator $A$ is unitary, the powers of $A$ will be also unitary. Let $U_m=A^m$, $U_n=A^n$, $ C_{m} =\| x(0)\|(\| \mathcal{M} \|t )^{m}/m!$, and $ D_{n}=\| b\|(\| \mathcal{M} \|t)^{n-1}t/{n!}$. By substituting them into Eq. (\ref{sim}), $\textbf{\emph{x}}(t)$ can be represented by
\begin{align}
\ket{x(t)} \approx  \frac{1}{\mathbb{N}^2}\big(\sum_{m=0}^{k}C_{m}U_{m}|x(0)\rangle  + \sum_{n=1}^{k}D_{n}U_{n-1}|b\rangle\big)
\label{eq10}
\end{align}
where $\mathbb{N}^2=\mathcal{C}^{2}+\mathcal{D}^{2}$ with $\mathcal{C}=\sqrt{\sum C_{m}}$ and  $\mathcal{D}= \sqrt{\sum D_{n}}$ is the normalization factor. Thus, the $j$-th element of $\textbf{\emph{x}}(t)$ would be $x_j(t)=\mathbb{N}^2\langle j|x(t)\rangle$.

\begin{figure}
\centering
\includegraphics[width=\columnwidth]{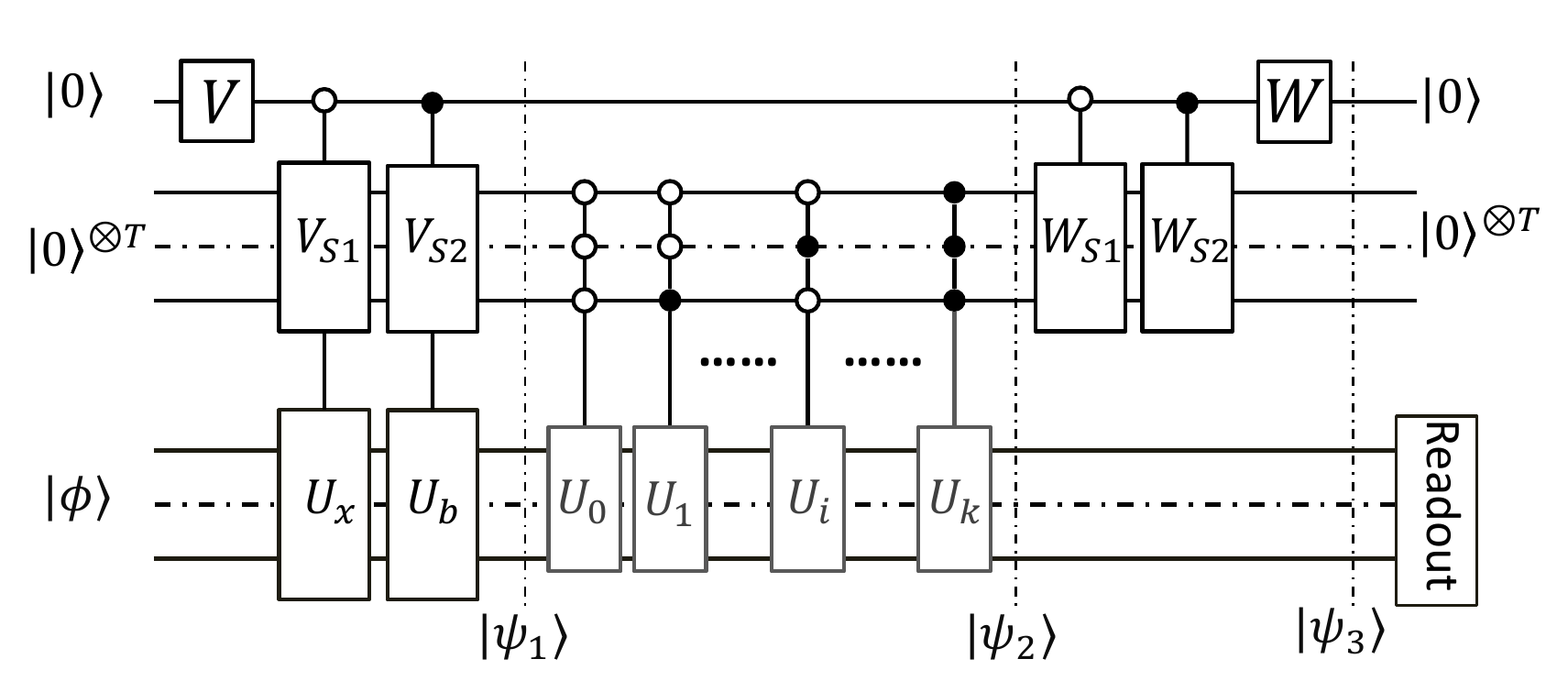}
\caption{ \footnotesize{Quantum circuit for solving LDEs when $A$ is unitary.  $\ket{\phi}$ denotes the initial state of the work system, and $T=\text{log}_{2}(k+1) $. The controlled operations $U_x$ and $U_b$ are used to create $\ket{x(0)}$ and $\ket{b}$, respectively.  The evolution operator during encoding and decoding is $ \sum_{\tau=0}^{k}|\tau\rangle \langle \tau|\otimes U_{\tau} $.  The state after each step is denoted as $|\psi_i\rangle, i=1,2,3$. At the end of the circuit, we measure the state vector of the work system in the subspace where all ancilla qubits are $\ket{0}$. }} \label{circuit_un}
\end{figure}

We employ a composite quantum system incorporating a work system and two ancilla registers to perform our algorithm as shown in Fig. \ref{circuit_un}. The  framework is divided into four steps as follows.

 (i) Encoding. $\text{log}_2N$ work qubits are needed to encode the $N$-dimensional vectors. $\ket{x(0)}$ and $\ket{b}$ are prepared and stored by the work qubits labeled by the subspace of the first ancilla register as $\ket{0}\ket{x(0)}$ and \ket{1}\ket{b}, respectively. In addition, a second ancilla register with $\text{log}_2k$ qubits is added and transformed into a specific superposition state $\ket{0}\sum_{m=0}^{k}\sqrt{ C_{m}}\ket{m}+\ket{1}\sum_{n=1}^{k}\sqrt{ D_{n}}\ket{n}$.

 Assume the input state of the work qubits is $|\phi\rangle$ and all ancilla qubits are $\ket{0}$ as shown in Fig. \ref{circuit_un}. The first operator $V$ is chosen as
 \begin{equation}
  V=\frac{1}{\mathbb{N}}\left(
  \begin{array}{cc}
    \mathcal{C} & \mathcal{D} \\
    \mathcal{D} &  -\mathcal{C}  \\
\end{array}
\right).
  \end{equation}
The encoded states $\ket{x(0)}$ and $\ket{b}$ are realized by performing controlled-operations $U_x$ and $U_b$ on the input state $\ket{\phi}$ depending on the state of the first ancilla qubit, respectively. A joint-controlled operation $\ket{0}\bra{0}\otimes V_{S1}\otimes U_x+\ket{1}\bra{1}\otimes V_{S2}\otimes U_b$ is applied subsequently, where $U_x$ and $U_b$ are used to evolve the work qubits into $\ket{x(0)}$ and $\ket{b}$, and $V_{S1}$ and $V_{S2}$ are $(k+1)\times (k+1)$ operations acting on the second ancilla register. The elements of the first rows in $V_{S1}$ and $V_{S2}$ are chosen as,
\begin{align}
V_{S1}(:,1)&=1/\mathcal{C} [\sqrt{C_{0}}, \sqrt{C_{1}}, ..., \sqrt{C_{k-1}}, \sqrt{C_{k}}],\\ \nonumber
V_{S2}(:,1)&=1/\mathcal{D} [\sqrt{D_{1}}, \sqrt{D_{2}}, ..., \sqrt{D_{k}}, 0],
\end{align}
while all other elements are arbitrary as long as  $V_{S1}$ and $V_{S2}$ are unitary. After computation, the initial state $|\psi_{in}\rangle = |0\rangle\otimes|0\rangle^{\otimes T}\otimes|\phi\rangle$ is evolved into:
\begin{align}
|\psi_{1}\rangle = \frac{1}{\mathbb{N}}\big(|0\rangle \sum_{m=0}^k\sqrt{C_m} |m\rangle \ket{x(0)} +
|1\rangle \sum_{n=1}^k\sqrt{D_n}|n-1\rangle \ket{b} \big).
\label{eq12}
\end{align}

 (ii) Entanglement creation.  A series of controlled operations are applied, to realize the operation $ \sum_{\tau=0}^{k}|\tau\rangle \langle \tau|\otimes U_{\tau} $ on the work qubits which is controlled by the second ancilla register. The work qubits and the ancilla registers are now entangled, and the whole state is
 \begin{align}
|\psi_{2}\rangle = & \frac{1}{\mathbb{N}}\big( |0\rangle \sum_{m=0}^k\sqrt{C_m}|m\rangle U_m\ket{x(0)}  \\\nonumber
&+ |1\rangle \sum_{n=1}^k\sqrt{D_n}|n-1\rangle U_{n-1}\ket{b}\big).\nonumber
\label{eq13}
\end{align}

 (iii) Decoding. All the operations in the encoding stage are reversely applied. $\ket{0}\bra{0}\otimes W_{S1}+\ket{1}\bra{1}\otimes W_{S2}$ on the ancilla registers are applied, where $W_{S1}=V_{S1}^{\dagger}$ and $W_{S2}=V_{S2}^{\dagger}$, followed by the last operator $W=V^{\dagger}$ on the first ancilla. Only the subspace where all ancilla qubits are $\ket{0}$ is concerned, and the state of the whole system in this subspace is
 \begin{equation}
|\psi_{3}\rangle = \frac{1}{\mathbb{N}^2}|0\rangle|0\rangle^{\otimes T}\big(\sum_{m=0}^{k}C_{m}U_{m}|x(0)\rangle  + \sum_{n=1}^{k}D_{n}U_{n-1}|b\rangle\big).
\label{eq15}
\end{equation}

 (iv) Measurement. Measure the final state of the work qubits in the subspace where all  ancilla qubits are $|0\rangle$.
It is obvious by comparing with Eq. (\ref{eq10}) that $\ket{x(t)}$ will be directly extracted, i.e., the solution to the LDE is obtained up to a factor $ \mathbb{N}^2$.

\textit{Case II}: This case that $A$ is non-unitary is similar to the first case, but more complicated. As $A$ can be decomposed into a linear combination of unitary operators $A=\sum\alpha_{i}A_{i}$ \cite{r1,r2,r3,Berry15,Wei16}, we need a third ancilla register to label the linear combinations $A_{i}$'s. Compared with the first case, we need more ancilla qubits and controlled operations. We leave details in the Supplemental Material~\cite{sm}.

\textit{Complexity. }--
Here we analyze the complexity of our algorithm for the case that $A$ is unitary (for the non-unitary case see the Supplemental Material~\cite{sm}).  The complexity involves two aspects: (1) Ancilla resources. As mentioned above, the number of total ancilla qubits is $1+ \text{log}_{2}(k+1)$. The order $k$ determines the gap $\epsilon$ between the ideal and approximate solutions by $k\leq  \ln(C_0/\epsilon)$ (proof in Appendix C \cite{sm}), where $C_0$ is a constant. (2) Query complexity. The successful rate of our method is roughly $1/s^2$, where $s$ is the amplitude of the state of the work qubits in the subspace $|0\rangle|0\rangle^{k}$ of the ancilla qubits. This rate can be improved by adopting the amplitude amplification by repeating the experiment $s$ times before measurement \cite{Berry15}. Hence, the total query complexity of our algorithm is about $O(sk)$ \cite{Berry15, Wei16}. More details can be found in the Supplemental Material \cite{sm}.

\textit{Experiment. }--
\begin{figure}
\centering
\includegraphics[width=\columnwidth]{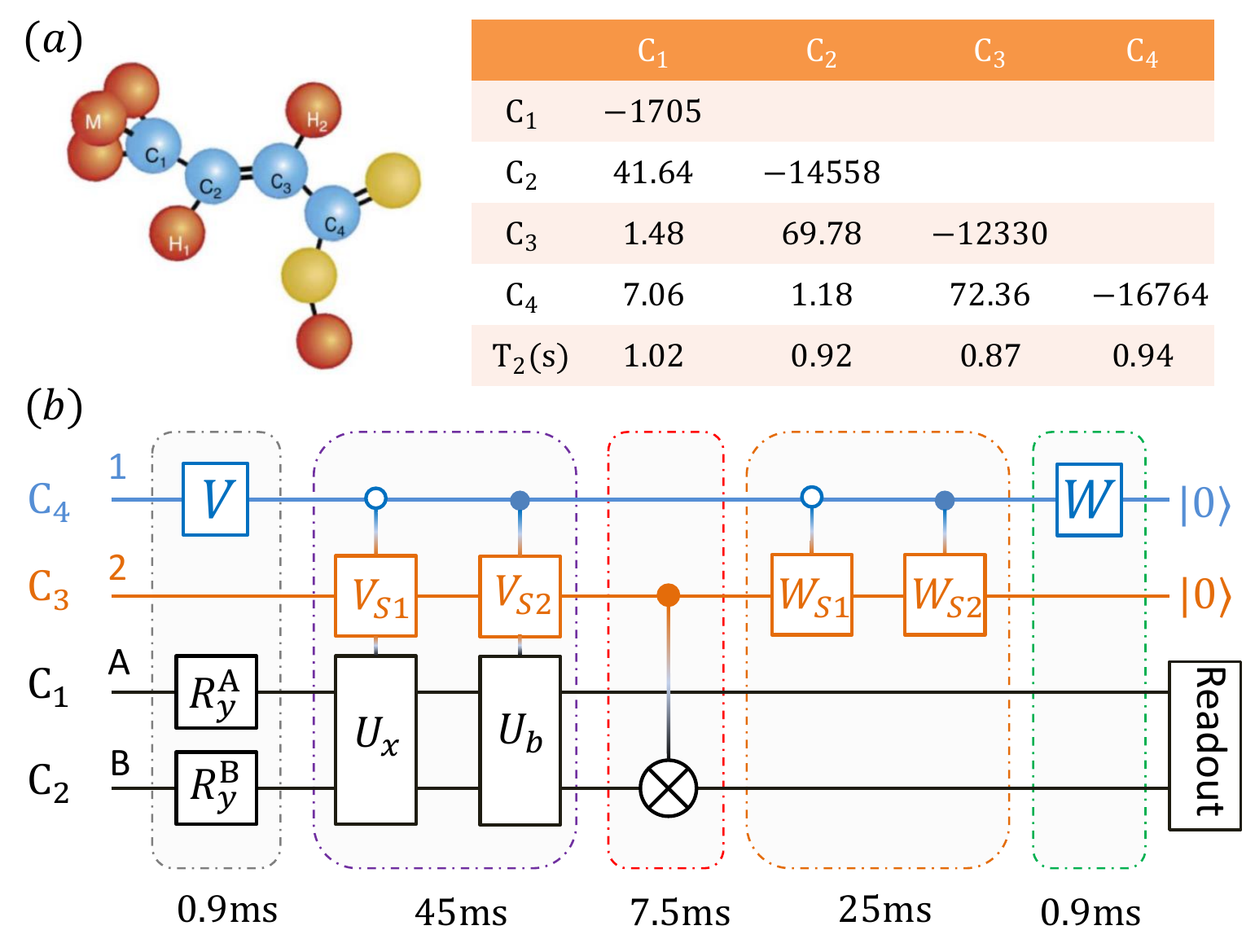}
\caption{ \footnotesize{(a) Molecular structure and Hamiltonian parameters of $^{13}$C-labeled trans-crotonic acid. C$_1$, C$_2$, C$_3$ and C$_4$ are used as four qubits in the experiment, while M, H$_1$ and H$_2$ are decoupled throughout the experiment. In the table, the chemical shifts with respect to the Larmor frequency and J-coupling constants (in Hz) are listed by the diagonal and off-diagonal numbers, respectively. The relaxation time scales T$_{2}$ (in seconds) are shown at the bottom. (b) NMR quantum circuit to realize the solution of a 4-dimensional LDE with four qubits}. A (labeled by C$_1$) and B (labeled by C$_2$) are work qubits to encode the vectors $\ket{x(0)}$ and $\ket{b}$.  Qubit 1 (labeled by C$_4$) and qubit 2 (labeled by C$_3$) are used as ancilla qubits. This circuit starts from $\ket{0000}$ which is prepared by the spatial average method. The input state $\ket{\phi}$ is then created by implementing the rotations $R_y^{\text{A}}(\beta_1)$ and $R_y^{\text{B}}(\beta_2)$ on the work qubits. $U_x=I\otimes I$ and $U_b=\sigma_x \otimes \sigma_x$ are applied to realize the preparation of the vectors $\ket{x(0)}$ and $\ket{b}$, respectively. Finally, we measure the state of the work qubits when the ancilla qubits are $\ket{00}$. Durations of the optimized pulses for each step are also given. } \label{expcircuit}
\end{figure}
In experiment, we demonstrate how to solve a 4-dimensional LDE with a $4\times 4$ non-unitary matrix  $\mathcal{M}$ ($A$ is thus also non-unitary). $\mathcal{M}$ is chosen as $\mathcal{M}=I\otimes I+2 I\otimes \sigma_x$, which can be decomposed into a linear combination of identity $\mathcal{M}_0=I\otimes I$ and pauli matrices $\mathcal{M}_1=I\otimes \sigma_x$. The initial vector $\ket{x(0)}$ and $\ket{b}$ are realized by applying two-qubit operations $U_x$ and $U_b$ on $\ket{\phi}$, respectively, where $\ket{\phi}$ is created using two single-qubit rotations on the input state $\ket{00}$. More specifically, $\ket{\phi}=R^\text{A}_y(\beta_1)R^\text{B}_y(\beta_2)\ket{00}$,
where $R^j_y(\beta)=e^{-i\beta\sigma^j_y/2}$ denotes a local rotation acting on qubit $j$ with angle $\beta$ about the $y$ axis.

The accuracy of the solution depends on the order $k$. We set the order $k=4$, leading to four qubits to implement the quantum circuit (see Fig. \ref{expcircuit}(c)) for solving the LDE. The forms of $V$, $W$, $U_c$, $V_{S1}$, $V_{S2}$, $W_{S1}$ and $W_{S2}$ can be found in the Supplemental Material \cite{sm}.
To experimentally realize our algorithm, we make use of the nuclear spins in a sample of $^{13}$C-labeled transcrotonic acid dissolved in d6-acetone \cite{nmr1,nmr2,nmr3}. The structure and parameters of the molecule are shown in Fig. \ref{expcircuit}(a).

\begin{table*}[tbp]
\centering
\begin{tabular}{|c|c|c|c|c|c|c|c|c|c|c|}
\hline
\hline
\multirow{2}*{$\beta_1$} & \multicolumn{2}{|c|}{0.1$\pi$}& \multicolumn{2}{|c|}{0.2$\pi$}& \multicolumn{2}{|c|}{0.3$\pi$}& \multicolumn{2}{|c|}{0.4$\pi$}& \multicolumn{2}{|c|}{0.5$\pi$} \\
\cline{2-11}
&theory & experiment &theory & experiment & theory & experiment & theory & experiment &theory & experiment\\
\hline
\multirow{4}*{Results $\textbf{\emph{x}}(t)$} &2.184 &2.136$\pm$0.017 &2.295&2.398$\pm$0.006 &2.305&2.276$\pm$0.004 &2.214&2.110$\pm$0.006 &2.030&1.916$\pm$0.007 \\
                                   &1.676 &1.570$\pm$0.046 &1.951&1.962$\pm$0.005 &2.110&2.186$\pm$0.004 &2.137&2.252$\pm$0.009 &2.030&2.176$\pm$0.011\\
                                   &0.635 &0.389$\pm$0.015 &1.066&0.804$\pm$0.004 &1.466&1.209$\pm$0.010 &1.799&1.525$\pm$0.011 &2.030&1.821$\pm$0.008\\
                                   &0.819 &0.693$\pm$0.016 &1.134&1.069$\pm$0.003 &1.462&1.482$\pm$0.003 &1.770&1.899$\pm$0.008&2.030&2.181$\pm$0.006\\
\hline
Similarity & \multicolumn{2}{|c|}{99.63\%$\pm$0.06\%}& \multicolumn{2}{|c|}{99.64\%$\pm$0.01\%}& \multicolumn{2}{|c|}{99.75\%$\pm$0.02\%}& \multicolumn{2}{|c|}{99.64\%$\pm$0.03\%}& \multicolumn{2}{|c|}{99.69\%$\pm$0.03\%}\\
\hline
\hline
\end{tabular}
\caption{\footnotesize{Experimental results of our algorithm for solving an LDE $d \textbf{\emph{x}}(t)/dt=\mathcal{M}\textbf{\emph{x}}(t)+\textbf{\emph{b}}$ at a fixed time $t=0.4$ s. } $\beta_1=\beta_2$ (see Fig. \ref{expcircuit}) ranges from $0.1\pi$ to $0.5\pi$ with a $0.1\pi$ increment. Theoretical and experimental solutions $\textbf{\emph{x}}(t)$ are both shown. To evaluate the performance, we compute the inner product (normalized) between the theoretical and experimental $\textbf{\emph{x}}(t)$. Error bars come from the uncertainty in repeated experiments, which is mainly attributed to the drift of temperature and inhomogeneity of the magnetic field. } \label{results}
\end{table*}

Firstly, we use the spatial averaging technique to prepare the pseudo-pure state (PPS) \cite{pps1,pps2,pps3} from the thermal equilibrium. The form of our 4-qubit PPS is $\rho_{0000}=(1-\epsilon)\mathbb{I}/16+\epsilon\ket{0000}\bra{0000}$, where $\mathbb{I}$ is the identity operator and the polarization $\epsilon\approx 10^{-5}$. Although the PPS is highly mixed, the large $\mathbb{I}$ does not evolve under unitary operations or contribute to the NMR signal. Hence, we only focus on the deviated part $\ket{0000}$. The fidelity between the ideal pure state $\ket{0000}$ and the experimental PPS is over 98\% by performing quantum state tomography \cite{sm}, which underpins subsequent experiments.

Subsequently, we perform the operations involved in our algorithm. All the operations are individually realized using shaped pulses optimized by the gradient method \cite{grape1,grape2,grape3}. Each shaped pulse is simulated to be over 0.995 fidelity while being robust to the static field distributions and inhomogeneity \cite{sm}.

Finally, we need to measure the state of the work qubits when the ancilla are $\ket{00}$. In experiment, we perform four-qubit state tomography to extract the desired results from the final density matrix \cite{qst1,qst2,sm}. It also enables us to evaluate the quality of  our implementation by comparing the distance between the target state $\rho_{th}$ and the experimentally reconstructed density matrix $\rho_{exp}$.
\begin{figure}
\centering
\includegraphics[width=\columnwidth]{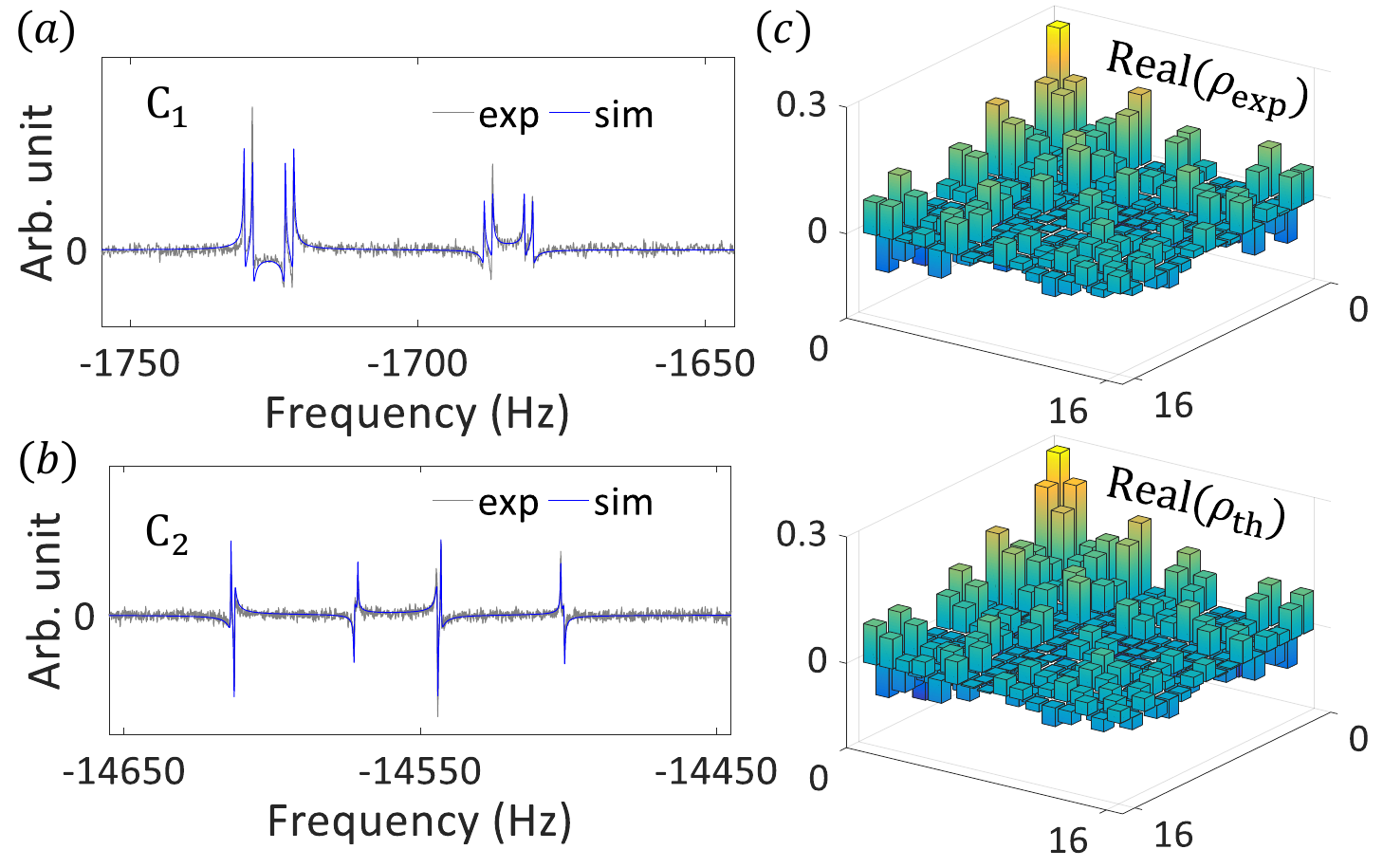}
\caption{ \footnotesize{ (a-b) NMR spectra of C$_1$ (work qubit A) and C$_2$ (work qubit B) followed by a readout pulse $R_x^{12}(\pi/2)R_x^{\text{AB}}(\pi/2)$ for $\beta = 0.1\pi$. The gray and blue lines show the experimental and simulated spectrum, respectively. (c) Real part of the density matrices $\rho_{exp}$ and $\rho_{th}$ for $\beta = 0.1\pi$. }} \label{spec}
\end{figure}

\textit{Results. }-- In experiment, we fix $t=0.4$ and $\beta_1=\beta_2$ ($\beta_1$ ranges from $0.1\pi$ to $0.5\pi$ with the increment $0.1\pi$). In other words, we demonstrate the solutions to five LDEs with different initial vectors $\ket{x(0)}$ and offset vectors $\ket{b}$ at a fixed time $t=0.4$.  For each value of $\beta_1$, the experiment is repeated by four times to estimate the uncertainty.  After the implementation of the quantum circuit, we perform the four-qubit state tomography by applying 17 readout pulses.
On average, the experimental fidelities for all states are about $0.936(5)$, estimated by
\begin{equation}
F(\rho_{th}, \rho_{exp})=\text{Tr}(\rho_{th}\rho_{exp})/\sqrt{\text{Tr}(\rho^2_{th})\text{Tr}(\rho^2_{exp})}.
\label{Fid}
\end{equation}
Taking $\beta=0.1\pi$ as an example, the comparison between the experimental and simulated NMR spectra of the work qubits is given in Fig. \ref{spec}(a-b), and they are in excellent agreement. The real parts of the density matrices for $\rho_{exp}$ and $\rho_{th}$ are also displayed in Fig. \ref{spec}(c) to evaluate the performance of our experiment. Furthermore, considering that $\mathcal{M}$, $\ket{x(0)}$, and $\ket{b}$ are all real in our setting, the solution $x(t)$ should be also real. We use a maximum likelihood (ML) approach  to construct a real state $\rho_{ml}$ which is closest to the experimental measured density matrix $\rho_{exp}$ \cite{mla1,nmr1}. After obtaining $\rho_{ml}$, we calculate the reduced state-vector of work qubits A and B in the subspace where the ancilla are $\ket{00}$, and then reproduce the solutions to the LDEs by amplifying the result by $\mathbb{N}^2=4.059$. Table \ref{results} summarizes all experimental results of the five LDEs and the comparison between theory and experiment.

The error in our experiment mainly comes from decoherence, imperfections of the input state preparation, and imprecisions of the optimized pulses. We numerically simulate each of the above factors to estimate the error distribution. For example, when $\beta_1=0.1\pi$, numerical results indicate that decoherence during the experimental running time 80 ms, the input state preparation, and imprecision of the optimized pulses lead to 1.5\%, 1.3\%, and 1.3\% infidelity, respectively. The sum of them (4.3\%) is slightly smaller than the total error $6.3\%$ in Eq. (\ref{Fid}). The additional 2\% error should be attributed to other error resources such as imperfection in the readout pulses and spectra fitting.

\textit{Conclusion. }--
We present a quantum algorithm and the relevant quantum circuit for solving LDEs with a precision $\epsilon$ by the number of ancilla resources growing as $ O(\text{ln} (C_0/\epsilon))$ and the number of queries growing as $O(sk)$. This precision naturally depends on the Taylor order $k$ and the number of ancilla qubits grows logarithmically with $k$. As a proof-of-principle demonstration, we experimentally realize the solution to a set of LDEs with the dimension $4\times 4$ in a four-qubit NMR quantum processor. The experimental solutions to these LDEs are obtained with about 6\% error, indicating the accuracy of the experimental implementation. We anticipate this algorithm to provide a key technique for many potential applications in the near future, such as route optimization of unmanned vehicles in artificial intelligence.

\begin{acknowledgments}
\textit{Acknowledgments. }--
T. X., S. W., and G. L. are grateful to the following funding sources: National Natural Science Foundation of China under Grants No. 11175094 and No. 91221205; National Basic Research Program of China under Grant No. 2015CB921002. I. A., L. L., and E. S. acknowledge financial support from Spanish MINECO/FEDER FIS2015-69983-P, Ram\'on y Cajal Grant RYC-2012-11391, and Basque Government IT986-16 and PhD grant PRE-2015-1-0394.

T. X. and S. W. contributed equally to this work. T. X. designed and performed the experimental scheme; S. W. proposed the theoretical approach; J. C. made the analysis for the error bounds; All the authors wrote and modified the paper.
\end{acknowledgments}

\clearpage

\onecolumngrid



\section{Supplemental Material for "A Quantum Algorithm for Solving Linear Differential Equation: Theory and Experiment"}

\subsection{Appendix A: Mathematical details of the algorithm}
We present a mathematical representation of our algorithm by considering the following two cases.

\textbf{A is unitary}:
In order to solve an LDE where matrix $A$ is unitary, we need a composite quantum system with a $(1+T)$-qubit ancilla register and a $\text{log}_2(N)$-qubit work system. Suppose the input state of the work system is $|\phi\rangle$ and all ancilla qubits are prepared in state $\ket{0}\ket{0}^{\otimes T}$. First, the first ancilla qubit evolves to a superposition state after  a unitary operation $V$ is performed,
 \begin{equation}
  V=\frac{1}{\mathbb{N}}\left(
  \begin{array}{cc}
    \mathcal{C} & \mathcal{D} \\
    \mathcal{D} &  -\mathcal{C}  \\
\end{array}
\right).
  \end{equation}
The encoded states $\ket{x(0)}$ and $\ket{b}$ are realized by performing controlled operations $U_x$ and $U_b$ on the input state $\ket{\phi}$, respectively. The initial state $|0\rangle|0\rangle^{\otimes T}|\phi\rangle$ is thus:
\begin{align}
\frac{\mathcal{C}}{\mathbb{N}}|0\rangle|0\rangle^{\otimes T}\ket{x(0)} +\frac{\mathcal{D}}{\mathbb{N}}|1\rangle|0\rangle^{\otimes T}\ket{b}.
\label{eq11}
\end{align}
Then, we define a $(k+1)\times (k+1)$ controlled operations $V_{S1}$ and $V_{S2}$ with
\begin{eqnarray}
 V_{S1}=\dfrac{1}{\mathcal{C}}\left(
  \begin{array}{cccccc}
  \sqrt{C_{0}}   & Q & Q& Q& Q & Q\\
  \sqrt{C_{1}}   & Q & Q& Q& Q & Q\\
 \cdots  & Q & Q& Q& Q & Q\\
 \sqrt{C_{k}}   & Q & Q& Q& Q & Q\\
\end{array}
\right)_{(k+1)\times (k+1)},\\
  V_{S2}=\dfrac{1}{\mathcal{D}}\left(
  \begin{array}{cccccc}
  \sqrt{D_{1}}   & Q & Q& Q& Q & Q\\
   \sqrt{D_{2}} & Q & Q& Q& Q & Q\\
 \cdots  & Q & Q& Q& Q & Q\\
\sqrt{D_{k}} & Q & Q& Q& Q & Q\\
    0  & Q & Q& Q& Q & Q\\
\end{array}
\right)_{(k+1)\times (k+1)}.
\end{eqnarray}
where $ Q 's$ are arbitrary elements that make $V_{S1}$ and $V_{S2}$ unitary. Then, Equation (\ref{eq11}) is changed to,
\begin{align}
 \frac{1}{\mathbb{N}}\left(|0\rangle \sum_{m=0}^k\sqrt{C_m} |m\rangle \ket{x(0)} +
|1\rangle \sum_{n=1}^k\sqrt{D_n}|n-1\rangle \ket{b} \right).
\label{eq12}
\end{align}
The controlled operation $U_c=\ket{0}\bra{0}\otimes U_0+\ket{1}\bra{1}\otimes U_1+...+\ket{k}\bra{k}\otimes U_k$ is implemented afterwards, where $U_k=A^k$. The state of the whole system is
 \begin{align}
\frac{1}{\mathbb{N}}\left( |0\rangle \sum_{m=0}^k\sqrt{C_m}|m\rangle U_m\ket{x(0)}
+ |1\rangle \sum_{n=1}^k\sqrt{D_n}|n-1\rangle U_{n-1}\ket{b}\right).
\label{eq13}
\end{align}
Subsequently, we implement the operations $W_{S1}=V_{S1}^{\dagger}$ and $W_{S2}=V_{S2}^{\dagger}$ controlled by the state of the first register on the second register, which leads to
 \begin{align}
\frac{1}{\mathbb{N}}\left( |0\rangle \ket{0}^{\otimes T} \sum_{m=0}^k\frac{C_m}{\mathcal{C}} U_m\ket{x(0)}
+ |1\rangle \ket{0}^{\otimes T} \sum_{n=1}^k\frac{D_n}{\mathcal{D}} U_{n-1}\ket{b}\right)
\label{eq14}
\end{align}
in the subspace where the second ancilla qubits are all  $\ket{0}^{\otimes T}$.
The last unitary operation $W=V^{\dagger}$ is applied on the first register. Analogously, we focus on the subspace where all ancilla qubits are $\ket{0}$, and the final state is
 \begin{equation}
\frac{1}{\mathbb{N}^2}|0\rangle|0\rangle^{\otimes T}\left(\sum_{m=0}^{k}C_{m}U_{m}|x(0)\rangle  + \sum_{n=1}^{k}D_{n}U_{n-1}|b\rangle\right).
\label{eq15}
\end{equation}
That is, if we measure the state of the work qubits in the subspace where ancilla are $\ket{0}\ket{0}^{\otimes T}$, the result directly represents the solution to the LDE amplified by a factor $ \mathbb{N}^{2}$. The successful probability of yielding the right answer is
\begin{eqnarray}
\frac{1}{(\mathbb{N}^{2})^{2}} \left(\sum_{m=0}^{k}C_{m}^{2}+\sum_{n=1}^{k}D_{n}^{2}\right) \approx \frac{1}{\mathbb{N}^{4}}.
\end{eqnarray}

\textbf{A is non-unitary}:
 First, the non-unitary matrix $A$ can be decomposed into a linear combination of unitary operators $A=\sum_{i=1}^{L}\alpha_{i}A_{i}$ where the $A_{i}$'s are unitary matrices. Thus, the  solution can be written as,
 \begin{eqnarray}
 |x(t)\rangle \approx \sum_{m=0}^{k}\dfrac{\| x(0)\|(\| \mathcal{M} \|  t)^{m}(\sum_{i=1}^{L}\alpha_{i}A_{i})^{m}}{m!}|x(0)\rangle  +
 \sum_{n=1}^{k}\dfrac{\| b \| \| \mathcal{M} \|^{n-1}t^{n}(\sum_{i=1}^{L}\alpha_{i}A_{i})^{n-1}}{n!}|b\rangle.
\end{eqnarray}
To obtain the solution, we need to add the third ancilla register compared to the case when $A$ is unitary. The first ancilla register is still encoded in one qubit. The second ancilla register consists of $k$ qubits, and the third ancilla register consists of $k$ qudits where each qudit is an $L$-level quantum system.

\begin{figure*}
\centering
\includegraphics[width=0.85\textwidth]{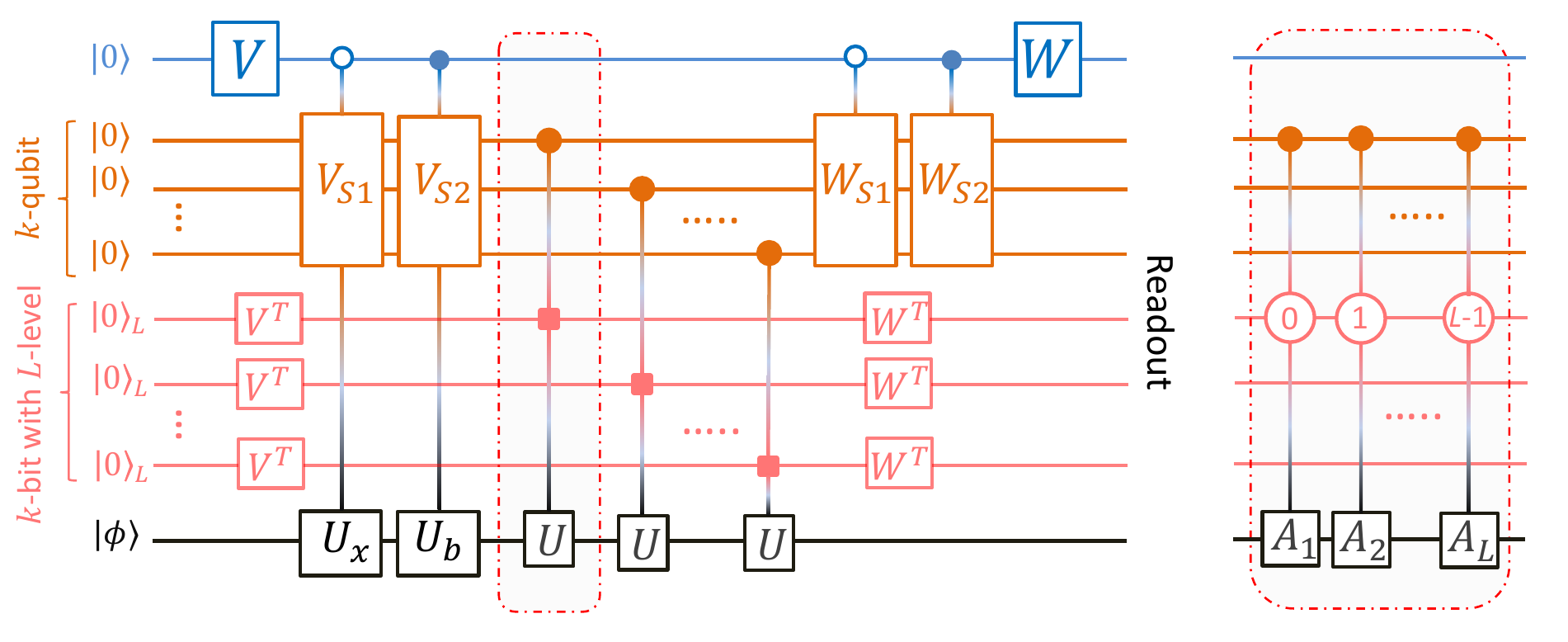}
\caption{  Universal quantum circuit for solving any LDEs.  The framework includes four parts: first ancilla register with one qubit, second ancilla register with $k$ qubits, third ancilla register with $k$ qudits where each qudit has $L$ levels and work system. All ancilla registers are initially prepared in the ground state $\ket{0}\ket{0}^k\ket{0}^k_L$. $\ket{0}_L$ denotes the ground state of an $L$-level system, which can be encoded by a log$_2(L)$-qubit quantum system. Hence, all operations acting on the third ancilla register are $L\otimes L$ matrices. The red squares denote jointly-controlled operations, with the corresponding circuit shown in the right.  At the end of the circuit, we measure the state of the work system in the subspace where  all ancilla registers are $\ket{0}\ket{0}^k\ket{0}^k_L$. } \label{circuit_any}
\end{figure*}

A universal quantum circuit to solve any LDE is illustrated in Fig. \ref{circuit_any}. Initially, all ancilla registers are prepared in the ground state $|0\rangle|0\rangle^{k}|0\rangle_{L}^{k}$, where $\ket{0}_L$ denotes the ground state of an $L$-level quantum system. The work system employs the input state $\ket{\phi}$ to subsequently encode the vectors $\ket{x(0)}$ and $\ket{b}$.
First, we implement the following operation $V$
on the first ancilla register,
\begin{equation}
  V=\left(
  \begin{array}{cc}
    \frac{G_{1}}{\sqrt{G_{1}^{2}+G_{2}^{2}}} & \frac{G_{2}}{\sqrt{G_{1}^{2}+G_{2}^{2}}} \\
     \frac{G_{2}}{\sqrt{G_{1}^{2}+G_{2}^{2}}} &  - \frac{G_{1}}{\sqrt{G_{1}^{2}+G_{2}^{2}}}  \\
\end{array}
\right).
  \end{equation}
where the parameters $G_1$ and $G_2$ are defined as
 \begin{eqnarray}
G_{1}=\sum_{m=0}^{k}\frac{\| x(0)\|(\| \mathcal{M} \|t )^{m}}{m!} (\sum_{i=1}^{L}\alpha_{i})^{m},
G_{2}=\sum_{n=1}^{k}\frac{\| b\|(\| \mathcal{M} \|t)^{n-1}t}{n!}(\sum_{i=1}^{L}\alpha_{i})^{n-1}.
\end{eqnarray}
In this way, we can encode the vectors $\ket{x(0)}$ and $\ket{b}$ by the controlled operations $U_x$ and $U_b$ on the work qubits, respectively.
We then perform the controlled operations $V_{S1}$ and $V_{S2}$ on the second ancilla register depending on the state of the first ancilla register. $V_{S1}$ and $V_{S2}$ are $2^k \times 2^k $ matrices. The $m$th element of the first column has the following definition,
\begin{align}
V_{S1}^{(m,1)}=\frac{v_{S1}^{(m,1)}}{\sqrt{\sum_{m} |v_{S1}^{(m,1)}|^2}},
V_{S2}^{(m,1)}=\frac{v_{S2}^{(m,1)}}{\sqrt{\sum_{m} |v_{S2}^{(m,1)}|^2}}
\end{align}
where
\begin{eqnarray}
v_{S1}^{(m,1)}=\left\{
\begin{array}{rcl}
\sqrt{\frac{\| x(0)\|(\| E \|t )^{j}}{j!}}, &  & {m=2^{k} -2^{k-j}+1},j\in \lbrace 0,1,\ldots ,k \rbrace. \\
 0   ,                             & & {\text{other case.}}
\end{array} \right.
\end{eqnarray}
\begin{eqnarray}
v_{S2}^{(m,1)}=\left\{
\begin{array}{rcl}
\sqrt{\frac{\| b\|(\| \mathcal{M} \|t)^{j-1}t}{j!}}, &  & {m=2^{k} -2^{k-j}+1},j\in \lbrace 1,2,\ldots ,k \rbrace. \\
 0   ,                             & & {\text{other case.}}
\end{array} \right.
\end{eqnarray}
Besides, we apply the unitary operation $V_T$ on each $L$-level qudit of the third ancilla register, where  $V_T$ is an $L\times L$ matrix. The $\ell$-th element of the first column in $V_T$ is
\begin{align}
V_{T}^{(\ell,1)}=\frac{v_{T}^{(\ell,1)}}{\sqrt{\sum_{\ell} |v_{T}^{(\ell,1)}|^2}}, \text{with} ~v_{T}^{(\ell,1)}=\sqrt{\alpha_{i}},
\end{align}
where $V^{T}_{\ell,0}=\sqrt{\alpha_{i}}$.
After implementing the unitary operations $V$, $V_{S1}$, $V_{S2}$ and $V_{T}$, the state of all ancilla registers will change from the initial state $|0\rangle|0\rangle^{k}|0\rangle_{L}^{k}$ to the following state,
\begin{eqnarray}
\frac{G_{1}}{\sqrt{G_{1}^{2}+G_{2}^{2}}} |0\rangle \sum_{m=0}^{2^k-1}V_{S1}^{(m+1,0)}|m\rangle \left(\sum_{\ell=1} ^{L}V_{T}^{(\ell,1)}|\ell-1 \rangle_{L}\right)^{\otimes k}\!\!\!\!+\frac{G_{2}}{\sqrt{G_{1}^{2}+G_{2}^{2}}} |1\rangle \sum_{m=0}^{2^k-1}V_{S2}^{(m+1,0)}|m\rangle \left(\sum_{\ell=1} ^{L}V_{T}^{(\ell,1)}|\ell-1 \rangle_{L}\right)^{\otimes k}\!\!\!\!\!\!\!\!.
\end{eqnarray}

To entangle the ancilla and the work qubits, we perform the controlled operation $U$ on the work system, which is jointly controlled by the states of the second and third ancilla registers. If we focus on the subspace $|0\rangle|0\rangle^{k}|0\rangle_{L}^{k}$ of all ancilla registers, the state of the work system can be written as,
\begin{align}
|x(0)\rangle+|b\rangle\rightarrow\frac{G_{1}}{\sqrt{G_{1}^{2}+G_{2}^{2}}} |0\rangle \sum_{m=2^{k} -2^{k-j}}V_{S1}^{(m+1,1)}|m\rangle \left(\sum_{\ell=1} ^{L}V_{T}^{(\ell,1)}A_{\ell}|\ell-1 \rangle_{L}\right)^{\otimes j}|x(0)\rangle\\
+\frac{G_{2}}{\sqrt{G_{1}^{2}+G_{2}^{2}}} |0\rangle \sum_{m=2^{k} -2^{k-j}}V_{S2}^{(m+1,1)}|m\rangle \left(\sum_{\ell=1} ^{L}V_{T}^{(\ell,1)}A_{\ell}|\ell-1 \rangle_{L}\right)^{\otimes j}|b\rangle.
\end{align}

For decoding, we need to perform the inverse operations on all ancilla registers.  The operations $W_{S1}=V_{S1}^{\dagger}$ and $W_{S2}=V_{S2}^{\dagger}$ are implemented on the second register, which is controlled by the state of the first register, and we reverse the the first and third registers by applying  $W=V^{\dagger}$ and $W_{T}=V_{T}^{\dagger}$, respectivey.  Finally, we measure the state of work qubits in the subspace where all ancilla registers stay on the state $|0\rangle|0\rangle^{k}|0\rangle_{L}^{k}$,
\begin{eqnarray}
|0\rangle|0\rangle^{k}|0\rangle_{L}^{k}|\phi\rangle &\rightarrow & \frac{1}{S}|0\rangle|0\rangle^{k}|0\rangle_{L}^{k} \left[\left(\sum_{m=0}^{k}\dfrac{\| x(0)\|(\| \mathcal{M} \|At)^{m}}{m!}\right)|x(0)\rangle + \left(\sum_{n=1}^{k}\dfrac{\| b \| (\| \mathcal{M} \| A)^{n-1}t^{n}}{n!}\right)|b\rangle\right].
 \label{eq9}
\end{eqnarray}
where $S= G_{1}+G_{2}$. One can obtain the solution to the LDE by multiplying $S$. If we directly measure the state of the work system at the end of circuit,  the probability of successfully detecting the auxiliary state $|0\rangle|0\rangle^{k}|0\rangle_{L}^{k} $ is
\begin{eqnarray}
P_{s}= \left|\left|\left[\left(\sum_{j=0}^{k}\dfrac{\| x(0)\|(\| \mathcal{M} \|At)^{j}}{j!}\right)|x(0)\rangle + \left(\sum_{j=1}^{k}\dfrac{\| b \| (\| \mathcal{M} \| A)^{j-1}t^{j}}{j!}\right)|b\rangle\right]\right|\right|/{S^{2}}
\end{eqnarray}
which is approximately $ 1/S^{2}$. Considering all the operations performed on the ancilla registers and work system, the total number of gates in our algorithm is $O(kL(\text{log}_{2}N + \text{log}_2L))$ \cite{Berry15sm}.

For this case, we can calculate the complexity of our algorithm in two aspects. (1) Ancillary resources. The number of total ancillary qubits is $1+ k+k\text{log}_{2}L$. The order $k$ determines the gap $\epsilon$ between the ideal and approximate solutions by the relationship $k\leq  \ln(C_0/\epsilon)$, with the constant $C_0$. (2)Query complexity. The successful probability of our method is roughly $1/S^2$, where $S$ is the amplitude of the state of work system on the subspace $|0\rangle|0\rangle^{k}|0\rangle_{L}^{k}$ of all registers. To improve the desired amplitude and obtain a near-100\% solution, we can adopt the robust obvious amplitude amplification by $S$ times before measurement. Hence, the total query complexity of our algorithm is about $O(kS)$.

\subsection{Appendix B: An Alternative approach of our algorithm}
When the matrix $A$ is non-unitary, we provide an alternative approach to solve LDEs. The solution can be written as,
 \begin{eqnarray}
 x(t) &\approx & \sum_{m=0}^{k}\dfrac{\| x(0)\|(\| \mathcal{M} \|At)^{m}}{m!}|x(0)\rangle  +\sum_{n=1}^{k}\dfrac{\| b \| (\| \mathcal{M} \| A)^{n-1}t^{n}}{n!}|b\rangle,
\end{eqnarray}
where $A$  is a normalized matrix satisfying $\|A\| \leq 1 $. Then, $A=B+iC $ where B and C are the real and imaginary parts with
\begin{eqnarray}
B=\frac{1}{2}(A+A^{\dagger}),\,\, C=\dfrac{1}{2i}(A-A^{\dagger}).
\end{eqnarray}
It is known that any real matrix can be decomposed into the linear combination of two unitary matrices. Hence,
\begin{eqnarray}
B=1/2 (F_1+F_2 ), \,\,C=1/2 (F_3+F_4 )
\end{eqnarray}
where the matrices $F_1, F_2$, $F_3$, and $F_4$ are all unitary. Their definitions are
\begin{eqnarray}
F_1=B+i\sqrt{I-B^{2}}, \,\,F_2=B-i\sqrt{I-B^{2}},\,\,
F_3=C+i\sqrt{I-C^{2}}, \,\,F_4=C-i\sqrt{I-C^{2}},
\end{eqnarray}
respectively. Then, we can obtain the relationship between the matrices $A$ and $F_i's$
 \begin{eqnarray}
 A=\dfrac{1}{2} (F_1+F_2 )+\frac{i}{2}(F_3+F_4 ).
 \end{eqnarray}
If the coefficient $i$ is absorbed into $F_3$  and $F_4$,
 \begin{eqnarray}
F_3=iC-\sqrt{I-C^{2}}, \,\,F_4=iC+\sqrt{I-C^{2}},
\end{eqnarray}
 we have
 \begin{eqnarray}
 A=\dfrac{1}{2} (F_1+F_2 +F_3+F_4 ).
 \end{eqnarray}
 The former is a linear combination of only four unitary matrices. In this situation, the solution $x(t)$ can be written as,
 \begin{eqnarray}
 \textbf{\emph{x}}(t) &\approx & \sum_{m=0}^{k}\dfrac{\| x(0)\|(\sum_{i=1}^{4}\frac{\| \mathcal{M} \|}{2}F_{i} t)^{m}}{m!}|x(0)\rangle  +\sum_{n=1}^{k}\dfrac{\| b \| (\sum_{i=1}^{4}\frac{\| \mathcal{M} \|}{2}F_{i})^{n-1}t^{n}}{n!}|b\rangle.
\end{eqnarray}
It shows that the number of unitary operators acting on $|x(0)\rangle$ is less than $  \frac{4^{k+1}-1}{3}$ and the number of unitary operators acting on $|b\rangle$ is less than $  \frac{4^{k}-1}{3}$. Thus, the number of required ancilla qubits is about log$_{2}( \frac{4^{k+1}-1}{3})+1 \approx 2k$.
The solution $|x(t)\rangle$ can be further expressed as
\begin{eqnarray}
 \textbf{\emph{x}}(t) &\approx & \sum_{m=1}^{\frac{4^{k+1}-1}{3}}C_{m}U_{m}|x(0)\rangle  +\sum_{n=1}^{\frac{4^{k}-1}{3}}D_{n}U_{n}|b\rangle.
\end{eqnarray}
The parameters $C_{m}$ and $D_{n}$ satisfy
\begin{equation}
C_{m} = \begin{cases}\| x(0)\|, &\textrm{$ m=1$,}\\
\frac{\| x(0)\|(\| \mathcal{M} \|t/2 )^{j}}{j!} & \textrm{log$_{4}(m-1) \leq j$, $ j\in {1,2,\cdots , k}$ }\end{cases}\label{eq:ce1}
\end{equation}
\begin{equation}
D_{n} = \begin{cases}\| b\|t, &\textrm{$ n=1$,}\\
\frac{\| b\|(\| \mathcal{M} \|t/2)^{j}t}{j!} & \textrm{log$_{4}(n-1) \leq j$, $ j\in {1,2,\cdots , k-1}$ }\end{cases}\label{eq:ce2}
\end{equation}
 Similarly, by defining $\mathcal{C}=\sqrt{\sum C_{m}}$ and $\mathcal{D}= \sqrt{\sum D_{n}}$, we obtain
 \begin{eqnarray}
 |x(t)\rangle \approx (\mathcal{C}^{2}+\mathcal{D}^{2})\left[\frac{1}{(\mathcal{C}^{2}+\mathcal{D}^{2})} \left(\mathcal{C}^{2} \frac{\left(\sum_{m=1}^{\frac{4^{k+1}-1}{3}}C_{m}U_{m}\right)}{\mathcal{C}^{2}}|x(0)\rangle + \mathcal{D}^{2}\frac{\left(\sum_{n=1}^{\frac{4^{k}-1}{3}}D_{n}U_{n}\right)}{\mathcal{D}^{2}}|b\rangle\right)\right].
\end{eqnarray}
This alternative approach provides a new way to realize the solution of any-type LDEs.

\subsection{Appendix C: Error bounds}

In this section, we analyze the infidelity between the exact solution $\tilde{\textbf{\emph{{x}}}}(t)$ and the approximate solution $\textbf{\emph{x}}(t)$, and give an upper bound of the error $ \epsilon= \parallel \textbf{\emph{x}}(t) -\tilde{\textbf{\emph{{x}}}}(t) \parallel$.
Since every square complex matrix is similar to a Jordan matrix, for an $n\times n$ complex matrix $\mathcal{M}$, there exists an $n\times n$ invertible matrix $T$ such that $\mathcal{M}=TJT^{-1}$, where $J=J_1\oplus J_2\oplus \cdots \oplus J_m$, and $J_i$ is a $d_i\times d_i$ Jordan block with eigenvalues $\lambda _i$,
\begin{equation*}
J_i=\left (\begin{array}{cccccc}
\lambda _i & 1 & 0 &\cdots & 0 & 0\\
0 & \lambda _i & 1 & \cdots & 0 & 0\\
0 & 0 & \lambda _i & \cdots & 0 & 0\\
\ldots & \ldots & \ldots & \ldots & \ldots & \ldots \\
0 & 0 & 0 &\cdots & \lambda _i & 1 \\
0 & 0 & 0 &\cdots & 0 & \lambda _i
\end{array}\right )
\end{equation*}
for $i=1, 2, \ldots m$, and $\sum _{i=1}^md_i=n$. Thus $e^{\mathcal{M}t}=Te^{Jt}T^{-1}.$
One can compute that
 $$e^{Jt}={\oplus }_{i=1}^me^{\lambda _it}J_i^\prime ,$$ where
\begin{equation*}
J_i^\prime =\left (\begin{array}{cccccc}
1& t & \frac{1}{2}t^2 &\cdots & \frac{1}{(d_i-2)!}t^{d_i-2}& \frac{1}{(d_i-1)!}t^{d_i-1}\\
0 & 1 & t & \cdots & \frac{1}{(d_i-3)!}t^{d_i-3} & \frac{1}{(d_i-2)!}t^{d_i-2}\\
0 & 0 & 1 & \cdots & \frac{1}{(d_i-4)!}t^{d_i-4} & \frac{1}{(d_i-3)!}t^{d_i-3}\\
\ldots & \ldots & \ldots & \ldots & \ldots & \ldots \\
0 & 0 & 0 &\cdots & 1 & t \\
0 & 0 & 0 &\cdots & 0 & 1
\end{array}\right )
\end{equation*}
is a $d_i\times d_i$ complex matrix.
It follows that
$$\|e^{Jt}\|=\max \{e^{t\mbox{Re}\lambda _i}\|J^\prime _i\|\mid i=1, 2, \ldots m\},$$
where $\|J^\prime _i\|$ denotes the spectral norm, that is, the largest singular value of $J_i^\prime$.
A Taylor expansion of $e^z$ with Lagrange remainder reads
$$e^z=\sum_{i=1}^k\frac{z^k}{k!}+\frac{e^{\theta z}}{(k+1)!}z^{k+1},$$
where $0<\theta <1$ is a constant.
Let $$C=\left(\|\textbf{\emph{x}}(0) \|+\frac{\|\textbf{\emph{b}}\|}{\|\mathcal{M}\|}\right)\|T\|\|T^{-1}\|\max\left\{e^{t\mbox{Re}\lambda _i }\|J^\prime _i\|\mid i=1, 2, \ldots ,m\right\}.$$
Then, the error is given by
$$\epsilon =\|\textbf{\emph{x}}(t) -\tilde{\textbf{\emph{{x}}}}(t) \|\leq \frac{\|\mathcal{M}t\|^{k+1}}{(k+1)!}C.$$
When $k$ is sufficiently large,
$$(k+1)!\approx\sqrt{2(k+1)\pi }\left(\frac{k+1}{e}\right)^{k+1},$$
it follows that
$$\frac{\sqrt{2\pi}}{C}\epsilon \leq \left(\frac{e\|\mathcal{M}t\|}{k+1}\right)^{k+1}\frac{1}{\sqrt{k+1}},$$ and hence,
\begin{equation*}
\mbox{ln}\frac{\sqrt{2\pi }}{C}\epsilon \leq (k+1)[\mbox{ln}\|e\mathcal{M}t\|-\mbox{ln}(k+1) ]-\frac{1}{2}\mbox{ln}(k+1)
\end{equation*}
so
\begin{equation*}
\mbox{ln}\frac{\sqrt{2\pi }}{C}\epsilon \leq (k+1)[\mbox{ln}\|e\mathcal{M}t\|-\mbox{ln}(k+1) ].
\end{equation*}
Since
$$\mbox{ln}(k+1) -\mbox{ln}\|e\mathcal{M}t\|\geq \frac{k+1-\|e\mathcal{M}t\|}{k+1},$$
 we have
 $$k+1\leq \|e\mathcal{M}t\|+\mbox{ln}\frac{C}{\sqrt{2\pi }}\frac{1}{\epsilon}.$$
Therefore,
$k\leq \mbox{ln}\frac{e^{\|e\mathcal{M}t\|-1}C}{\sqrt{2\pi }}\frac{1}{\epsilon }$. Let $C_0=\frac{e^{\|e\mathcal{M}t\|-1}C}{\sqrt{2\pi }}$. Then, $k\leq \mbox{ln}\frac{C_0}{\epsilon },$
which implies that the larger the value of $k$, the smaller the error $\epsilon$.

\subsection{Appendix D: Experimental molecule and PPS preparation}

Experimentally, we demonstrate the quantum algorithm for solving a 4-dimensional LDE with a four-qubit nuclear magnetic resonance system. We make use of the nuclear spins in a sample of $^{13}$C-labeled transcrotonic acid dissolved in d6-acetone. The internal Hamiltonian of this system can be described as
 \begin{align}
\mathcal{H}_{int}=\sum\limits_{j=1}^4 {\pi \nu _j } \sigma_z^j  + \sum\limits_{j < k,=1}^4 {\frac{\pi}{2}} J_{jk} \sigma_z^j \sigma_z^k.
\end{align}
where $\nu_j$ is the chemical shift of the $\emph{j}$th spin and $\emph{J}_{jk}$ is the J-coupling strength between spins $\emph{j}$ and $\emph{k}$. We assigned C$_1$ and C$_2$ as system qubits, and C$_4$ and C$_3$ as ancilla qubits, respectively. All experiments were carried out on a Bruker ADVANCE 400 MHz spectrometer at room temperature.

\begin{figure*}
\centering
\includegraphics[width=0.8\textwidth]{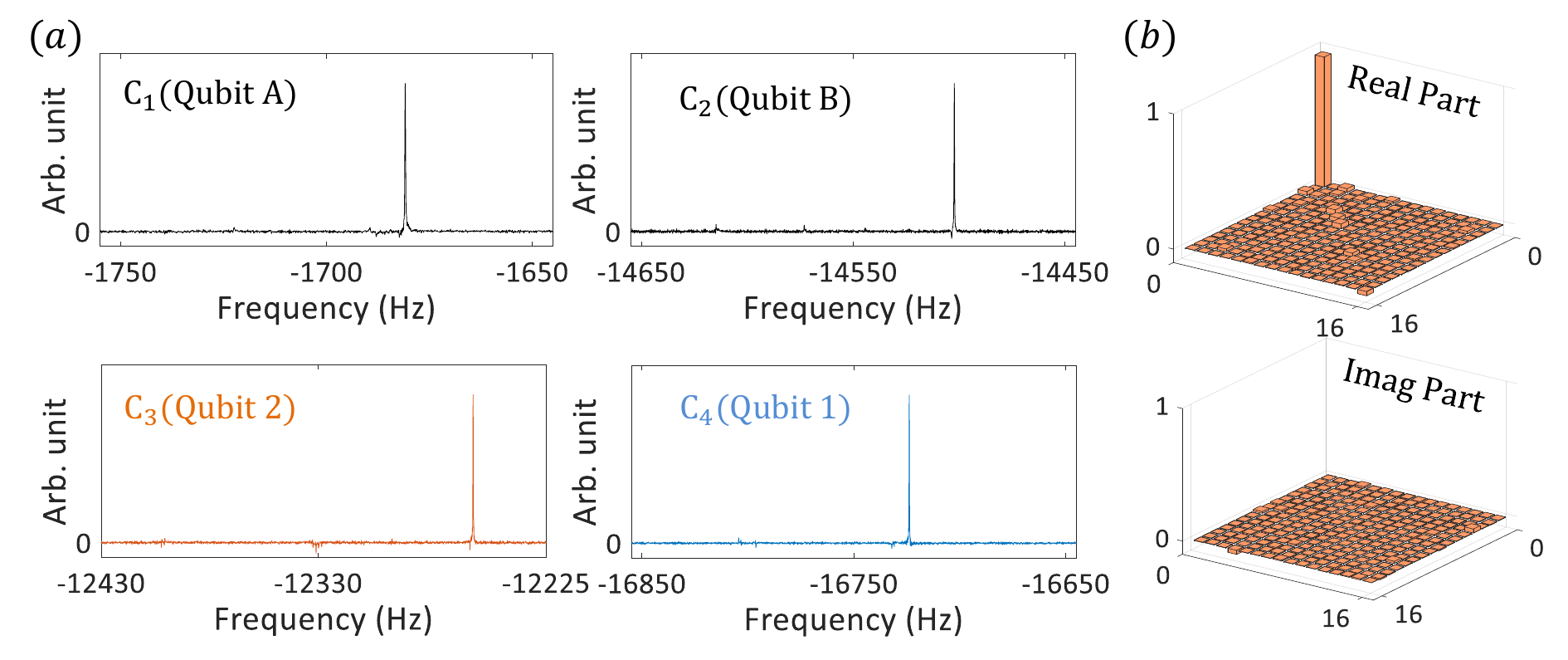}
\caption{ \footnotesize{Experimental spectra of the nuclei C$_1$ to C$_4$ and the reconstructed density matrix of the PPS}. (a) NMR signals of the nuclei C$_1$ to C$_4$ are measured by applying the corresponding $\pi/2$ readout pulses after the PPS preparation.  (b) Top and bottom plots respectively show the real and imaginary part of the reconstructed PPS matrix. The $z$ axis represents the value of the element in the matrix.  } \label{exp_pps}
\end{figure*}

At thermal equilibrium, an NMR sample stays in the Boltzmann distribution,
\begin{align}\label{thermal}
\rho_{\text{thermal}}=\frac{{\mathcal{I}}}{16}+\epsilon(\sigma_z^1+\sigma_z^2+\sigma_z^3+\sigma_z^4),
\end{align}
where $\mathcal{I}$ is a $16\times 16$ identity matrix and the polarization $\epsilon\approx10^{-5}$.  It is a highly-mixed state which is not suitable for quantum computing. Starting from this state, we use the spatial averaging technique to realize the preparation of the following PPS,
\begin{align}\label{pps}
\rho_{0000}=\frac{1-\epsilon}{16}{\mathcal{I}}+\epsilon\ket{0000}\bra{0000}.
\end{align}
The initialization processing usually includes local unitary rotations and $z$-gradient fields for supressing the undesired coherence.  Considering that the identity part does not evolve under any unitary operations or influences any measurements in NMR, the deviation density matrix $\ket{0000}\bra{0000}$ can serve as the initial state of the quantum circuit. Figure \ref{exp_pps} presents experimental spectra of the PPS for different carbon nuclei and the reconstructed density matrix of the PPS by performing state tomography.

\section{Appendix E: Experimental protocol}

In experiment, the parameters of the target LDE are chosen as follows. $\mathcal{M}$ is chosen as $\mathcal{M}=I\otimes I+2 I\otimes \sigma_x$. Starting from the initial state $\ket{\phi}$, we encode the vector $\ket{x(0)}$ by applying a two-qubit operation $U_x$ on $\ket{\phi}$, and the offset vector $\ket{b}$ by applying an additional rotation $U_b$ on $\ket{\phi}$. More specifically,
\begin{align}
\ket{\phi}=R^\text{A}_y(\beta_1)R^\text{B}_y(\beta_2)\ket{00}, \,\,
\ket{x(0)}=U_x\ket{\phi}, \,\,\ket{b}=U_b\ket{\phi}.
\label{eq16}
\end{align}
where $R^j_y(\beta)$ denotes a local rotation $R^j_y(\beta) =e^{-i\beta\sigma^j_y/2}$ acting on qubit $j$ with angle $\beta$ about the $y$-axis. The order $k$ in the Taylor expansion directly determines the accuracy of the approximate solution $\textbf{\emph{x}}(t)$. We choose the order $k=4$. The corresponding solution is,
\begin{align}
\textbf{\emph{x}}(t)\approx \left[\left(1+t+\frac{5t^2}{2}+\frac{13t^3}{6}+\frac{41t^4}{24}\right)U_0
+\left(2t+2t^2+\frac{7t^3}{3}+\frac{5t^4}{3}\right)U_1\right]\ket{x(0)} +\\
\left[\left(t+\frac{t^2}{2}+\frac{5t^3}{6}+\frac{13t^4}{24}\right)U_0+\left(t^2+\frac{2^2}{3}+\frac{7t^4}{12}\right)U_1\right]\ket{b}.\nonumber
\label{eq1170}
\end{align}
As shown in Fig. 2 of the main text, we present a detailed quantum circuit with four qubits for realizing the solution $\textbf{\emph{x}}(t)$. $\mathcal{C}$ and $\mathcal{D}$ in operations $V$ and $W$ are defined as
\begin{align}
\mathcal{C}&= \sqrt{\left(1+t+\frac{5t^2}{2}+\frac{13t^3}{6}+\frac{41t^4}{24}\right)+\left(2t+2t^2+\frac{7t^3}{3}+\frac{5t^4}{3}\right)}\\\nonumber
\mathcal{D}&=\sqrt{\left(t+\frac{t^2}{2}+\frac{5t^3}{6}+\frac{13t^4}{24}\right)+\left(t^2+\frac{2^2}{3}+\frac{7t^4}{12}\right)}.
\label{eq17}
\end{align}
Operations $V_{S1}$ and $V_{S2}$ are chosen as
\begin{eqnarray}
 V_{S1}=\dfrac{1}{\mathcal{C}}\left(
  \begin{array}{cc}
  \sqrt{1+t+\frac{5t^2}{2}+\frac{13t^3}{6}+\frac{41t^4}{24}}   & N \\
  \sqrt{2t+2t^2+\frac{7t^3}{3}+\frac{5t^4}{3}}    & N\\
\end{array}
\right),\,\,
  V_{S2}=\dfrac{1}{\mathcal{D}}\left(
  \begin{array}{cc}
  \sqrt{t+\frac{t^2}{2}+\frac{5t^3}{6}+\frac{13t^4}{24}}   & N\\
   \sqrt{t^2+\frac{2^2}{3}+\frac{7t^4}{12}} & N\\
\end{array}
\right).
\end{eqnarray}
$ N 's$ are arbitrary elements that make $V_{S1}$ and $V_{S2}$ unitary, which can be determined using the Gram-Schmidt method. The other operations are $W_{S1}=V_{S1}^{\dagger}$ and $W_{S2}=V_{S2}^{\dagger}$.  The controlled operation $U_c$ is simplified to the controlled-NOT operation $U_c=I\otimes (\ket{0}\bra{0}\otimes I+\ket{1}\bra{1}\otimes \sigma_x) \otimes I$.

\end{document}